\documentclass[aps,pra,twocolumn,superscriptaddress,showpacs]{revtex4-1}

\usepackage{amsmath,amssymb,amsfonts}
\usepackage{graphicx}
\usepackage{subfigure}
\usepackage{hyperref}

\newcommand{\Op}[1]{\hat{#1}}
\newcommand{\Opd}[1]{\hat{#1}^\dagger}
\newcommand{\op}[2]{\hat{#1}_{#2}}
\newcommand{\opd}[2]{\hat{#1}_{#2}^\dagger}

\newcommand{\trans}{\mathsf{T}}
\newcommand{\fref}[1]{Fig.~\ref{#1}}

\newcommand{\Tr}{\mathrm{Tr}}
\newcommand{\sig}{\boldsymbol{\sigma}}

\begin{document}

\title{Critical behavior in ultra-strong-coupled oscillators}

\author{Vivishek Sudhir}
\altaffiliation[Current address: ]{LPQM, \'{E}cole Polytechnique F\'{e}d\'{e}rale de Lausanne (EPFL), 1015 Lausanne, Switzerland}
\affiliation{QOLS, Blackett Laboratory, Imperial College London, London SW7 2BW, UK}

\author{Marco G. Genoni}
\affiliation{QOLS, Blackett Laboratory, Imperial College London, London SW7 2BW, UK}

\author{Jinhyoung Lee}
\affiliation{Physics Department, Hanyang University, Sungdong-Gu, Seoul 133-791, South Korea}

\author{M. S. Kim}
\affiliation{QOLS, Blackett Laboratory, Imperial College London, London SW7 2BW, UK}

\date{\today}

\begin{abstract}
We investigate the strong coupling regime of a linear $x$-$x$ coupled harmonic oscillator system, by performing a direct diagonalization
of the hamiltonian. It is shown that the $x$-$x$ coupled hamiltonian can be equivalently described by a Mach-Zehnder-type interferometer
with a quadratic unitary operation in each of its arms. We show a sharp transition of the unitary operation from an elliptical phase
rotator to an elliptical squeezer as the
coupling gets stronger, which leads to the continuous generation of entanglement, even for a significantly thermal state,
in the ultra-strong coupled regime. It is also shown that this
critical regime cannot be achieved by a classical Hookian coupling. Finally, the effect of a finite-temperature environment
is analyzed, showing that entanglement can still be generated from a thermal state in the ultra-strong coupled regime, but 
is destroyed rapidly.
\end{abstract}
\maketitle

\section{Introduction}
Strong coupling, which is one of the key
ingredients for the manipulation and control of quantum systems, makes quantum-mechanical predictions distinct from classical ones. 
In fact, the definition of strong coupling
in quantum mechanics has evolved over the
years. In atomic and optical physics, strong coupling was defined to show Autler-Townes splitting
\cite{AutTow55} where atom-field interaction is stronger than the atomic decay rate. The next age of strong coupling was defined 
in cavity quantum electrodynamics (QED), where an atom-field interaction strong in comparison to
the cavity and atomic decay rates can be realized. A series of important experiments have been performed to show the nonclassical 
nature of the atom-field interaction for Rydberg atoms in microwave cavities \cite{RaiHar01,HarBook}. 
This regime was recently also attained with a cavity optomechanical system \cite{Verh12}. 
But with a system consisting of superconducting Josephson junctions coupled to microwave fields in stripline
resonators, the interaction strength can be made even stronger than the atom-field coupling realized previously in 
cavity QED \cite{WalScho04,NieSol10}.
These so-called circuit QED systems can
reach coupling strengths comparable to the transition frequency of the relevant qubit system. In these systems,
we find new phenomena such as the breakdown of the rotating-wave approximation and the emergence of a unique deep strong coupling
regime \cite{CasSol10}.

A system of coupled oscillators has been of interest in various contexts, as several physical systems are represented by 
harmonic oscillators
such as nanomechanical oscillators, electromagnetic fields and a bunch of two-level systems.
Hopfield studied the quantization of an electromagnetic field in a dispersive medium using a coupled harmonic oscillator model \cite{Hop58}.
Recently, such a system of coupled oscillators has been
considered as a possible mechanism to generate entanglement in an array of nanoelectromechanical 
devices \cite{EisPlenBos04,AudPlenWer02,Galv10}.
In this paper, we study ultra-strong coupling between
harmonic oscillators where the coupling rate $g$ is comparable to or larger than the natural frequencies $\omega_j$ of the
harmonic oscillators.
Strongly coupled oscillators, where the counter-rotating terms are not negligible have been studied in the context of twin-photon
generation \cite{CiuCar05}.
The study of such systems are supported by recent experimental progress wherein such a regime is becoming accessible,
for instance in the interaction of multiple quantum well structures with terahertz electromagnetic fields \cite{GunCiuHub05}. In this paper,
we go even further than the strong
coupling regime, and find a critical point in the coupling strength
at which the evolution of the oscillator states radically changes. An equivalent regime was studied for the interaction between an atom 
and a field \cite{CasSol10,bina11}, where photon
number wavepackets were found to experience collapse and revivals across parity chains in Hilbert space.
The model that we consider here, is quite different in the sense that we have a bosonic system living in an infinite dimensional
Hilbert space, so that there is no such parity-defined dynamics. However, the model is exactly solvable and we find critical 
effects from the dynamics of the coupled oscillators, which get manifested in their entanglement behaviour.

We consider a system of two bosonic oscillators interacting via a quadratic interaction, whose hamiltonian is described by
\begin{equation}\label{hamiltonian}
    \Op{H} = \sum_{j=1}^2\frac{\omega_j}{2}(\op{p}{j}^2 +\op{x}{j}^2) + g \op{x}{1}\op{x}{2},
\end{equation}
where the dimensionless quadrature operators satisfy the canonical commutation relation, i.e. $[\op{x}{j},\op{p}{k}] =i\delta_{jk}$.
For instance, the Dicke model of superradiance \cite{Dicke68} is approximately described by such a hamiltonian.
In the context of cavity optomechanical systems, such a linearized hamiltonian describes the dynamics when the mechanics is 
being driven by a
strong coherent field \cite{MancTomb94}. In fact, this linearized model has been investigated previously in the optomechanics context
\cite{AkrAspMil10} in the rotating wave approximation (RWA), which is possible when the system is relatively weakly coupled.

In this paper, we will show that in the ultra-strong coupling regime, the non-RWA terms turn out to be not only important but also to
critically change the oscillator interaction. After identifying a unitary transformation
exactly diagonalizing the hamiltonian in \eqref{hamiltonian}, we show its equivalence to a Mach-Zehnder-type setup with 
unitary operations in its
arms, which explains the generation of entanglement in the model. At the critical point when the coupling becomes ultra-strong, 
we observe the transition of the unitary operation from an elliptic rotator to a squeezer. This transition is clearly reflected 
in the entanglement dynamics of the oscillators. It is interesting to note that the classical Hookian model for classical harmonic
oscillators is not able to reach this critical point.

\section{Diagonalization of the hamiltonian}
One can codify the operator hamiltonian \eqref{hamiltonian} in terms of a hamiltonian matrix,
\begin{equation}\label{hamiltonian_matrix}
    H = \begin{bmatrix}
          \omega_1 & g \\
          g & \omega_2 \\
        \end{bmatrix} \oplus \begin{bmatrix}
                               \omega_1 & 0 \\
                               0 & \omega_2 \\
                             \end{bmatrix} \equiv H_x \oplus H_p,
\end{equation}
expressed in the basis defined by the column vector of quadratures $\Op{R}=(\op{x}{1},\op{x}{2},\op{p}{1},\op{p}{2})^\trans$, so that
$\Op{H}=\frac{1}{2}\Op{R}^\trans H\Op{R}$ where we also define the diagonal block matrices corresponding to positions and momenta.

The usual diagonalization of such a quadratic hamiltonian proceeds by defining normal mode operators $\op{\Omega}{q}$ ($q=\pm$) 
and their hermitian conjugates, which are linear combinations of the bare quadratures satisfying $[\op{\Omega}{q},\Op{H}]=E_q 
\op{\Omega}{q}$.  Specifically, the normal mode energies are found to be $E_\pm^2 = \omega_1^2 +\omega_2^2 \pm \sqrt{ (\omega_1^2 
+\omega_2^2)^2 + 4\omega_1\omega_2(g^2 -\omega_1 \omega_2)}$, from which we already observe an interesting regime where $E_-$ is 
imaginary, for $g>g_c \equiv \sqrt{\omega_1 \omega_2}$.

The transformation to normal modes is just a symplectomorphism \cite{DufSal98}.
The hamiltonian matrix can be diagonalized by a symplectic matrix $S \in Sp(4,\mathbb{R})$ such
that it is block diagonal of the form $S=S_x \oplus S_p$. The diagonal hamiltonian matrix is then $H' = S^\trans H S =
(S_x^\trans H_x S_x)\oplus (S_p^\trans H_p S_p)$. From the general conditions to be satisfied by symplectic transformations, we get that for
$S_x, S_p \in SL(2,\mathbb{R})$, $S_p^\trans = S_x^{-1}$. The unitary representation of $S$ in the Hilbert space of the system is,
\begin{equation} \label{T-op}
    \Op{T}=\exp\left[i (A \op{x}{1}\op{p}{2}- B\op{x}{2}\op{p}{1}) \right],
\end{equation}
parameterized by real constants $A$ and $B$. This diagonalizes $\Op{H}$ to $\Op{H}'=\Op{T}\Op{H}\Opd{T}$ iff.
\begin{equation}
    \tan 2\sqrt{AB} = \frac{2gg_c}{\omega_1^2 -\omega_2^2} \text{  and  } \frac{A}{B}=\frac{\omega_2}{\omega_1}.
\label{condition}
\end{equation}
The diagonal hamiltonian is then given by,
\begin{eqnarray}\label{d-hamiltonian}
    \Op{H}' = & \omega_1\op{p}{1}^2 + \left(\frac{\omega_1}{2}c^2 +\frac{\omega_2^2}{2\omega_1}s^2 
            +\frac{gg_c}{\omega_1}cs \right)\op{x}{1}^2 \nonumber \\
            &+ \omega_2\op{p}{2}^2 + \left(\frac{\omega_2}{2}c^2 +\frac{\omega_1^2}{2\omega_2}s^2 
            -\frac{gg_c}{\omega_2}cs \right)\op{x}{2}^2,
\end{eqnarray}
where we denote $c=\cos \sqrt{AB}$ and $s=\sin \sqrt{AB}$. Each normal mode is described by a quadratic hamiltonian,
which looks very similar to a standard free hamiltonian for an oscillator. But using the condition (\ref{condition}), it is found that the
last term in \eqref{d-hamiltonian} is zero (negative) when $g=g_c$ ($g>g_c$), so that these are standard harmonic oscillators 
only when $g<g_c$.
The first (second) mode in \eqref{d-hamiltonian} is associated with $E_+$ ($E_-$), which we call the `$+$' (`$-$') mode.
When $g=g_c$, the `$-$' mode does not have a bound spectrum, but is rather a free particle, while the `$+$' mode is still a 
harmonic oscillator. Increasing the coupling strength further, for $g>g_c$, the anomalous
`$-$' mode is dynamically unstable since it is driven by a force derived from the inverted harmonic potential. So, the 
earlier observation of $E_-$ going imaginary is reflected by this qualitative change in the dynamical behaviour of the 
system, as the coupling strength crosses its critical value $g_c$.

\section{Inequivalence to classical coupled oscillators}
An obvious intuition one would have from thinking about the classical oscillating systems is that the model \eqref{hamiltonian} must be
the quantized version of the classical Hookian hamiltonian for coupled oscillators,
\begin{equation}\label{hamiltonian_classical}
    \Op{H}_\text{C} = \sum_{j=1}^2\left( \frac{\op{P}{j}^2}{2m} + \frac{1}{2}m\omega^2 \op{X}{j}^2 \right)
                    + \frac{m{\cal G}^2}{2} \left(\op{X}{1}-\op{X}{2}\right)^2,
\end{equation}
where $m$ is the common mass of the particle, $\omega$ the common natural frequency and ${\cal G}$ the Hookian coupling rate. 
The oscillators are assumed to be identical for simplicity (a generalization is straightforward). Introducing
dimensionless quadratures, $\op{x}{1,2} = (m\omega_0)^{1/2} \op{X}{1,2}, \,\,\, \op{p}{1,2} = (m\omega_0)^{-1/2} \op{P}{1,2}$ 
with the renormalized frequency $\omega_0 = \sqrt{\omega^2 + {\cal G}^2}$, the hamiltonian becomes,
\begin{equation*}
    \op{H}{\text{C}} = \frac{\omega_0}{2} \left[ (\op{p}{1}^2 + \op{x}{1}^2) + (\op{p}{2}^2 + \op{x}{2}^2) - G \op{x}{1}\op{x}{2} \right],
\end{equation*}
where $G=\frac{{\cal G}^2}{\omega^2 + {\cal G}^2}$ is the renormalized coupling strength. So, it is clear that in the quantum version of the
classical Hookian problem, the ultra strong coupling regime (here, $G>1$) is not possible to achieve.

The ultra strong coupling transition of the form exhibited by the hamiltonian \eqref{hamiltonian} seems to be a property of 
such quantum coupling models, and not something that is seen easily in the standard harmonic oscillator model. In particular 
though, for optomechanical problems, the linearization of the actual
radiation pressure interaction $(\op{p}{1}^2 +\op{x}{1}^2)\op{x}{2}$ (which leads to $\op{x}{1}\op{x}{2}$) is valid only for a short
interaction time. Otherwise, the displacement of the mechanics gets too large and the approximation leading to the interaction hamiltonian
breaks down in the strong coupling regime $g>g_c$ \cite{Law94,CollWalls93,VanPNAS}.

\section{Entanglement dynamics}
\begin{figure}[h!]
    \centering
    \includegraphics[scale=0.4]{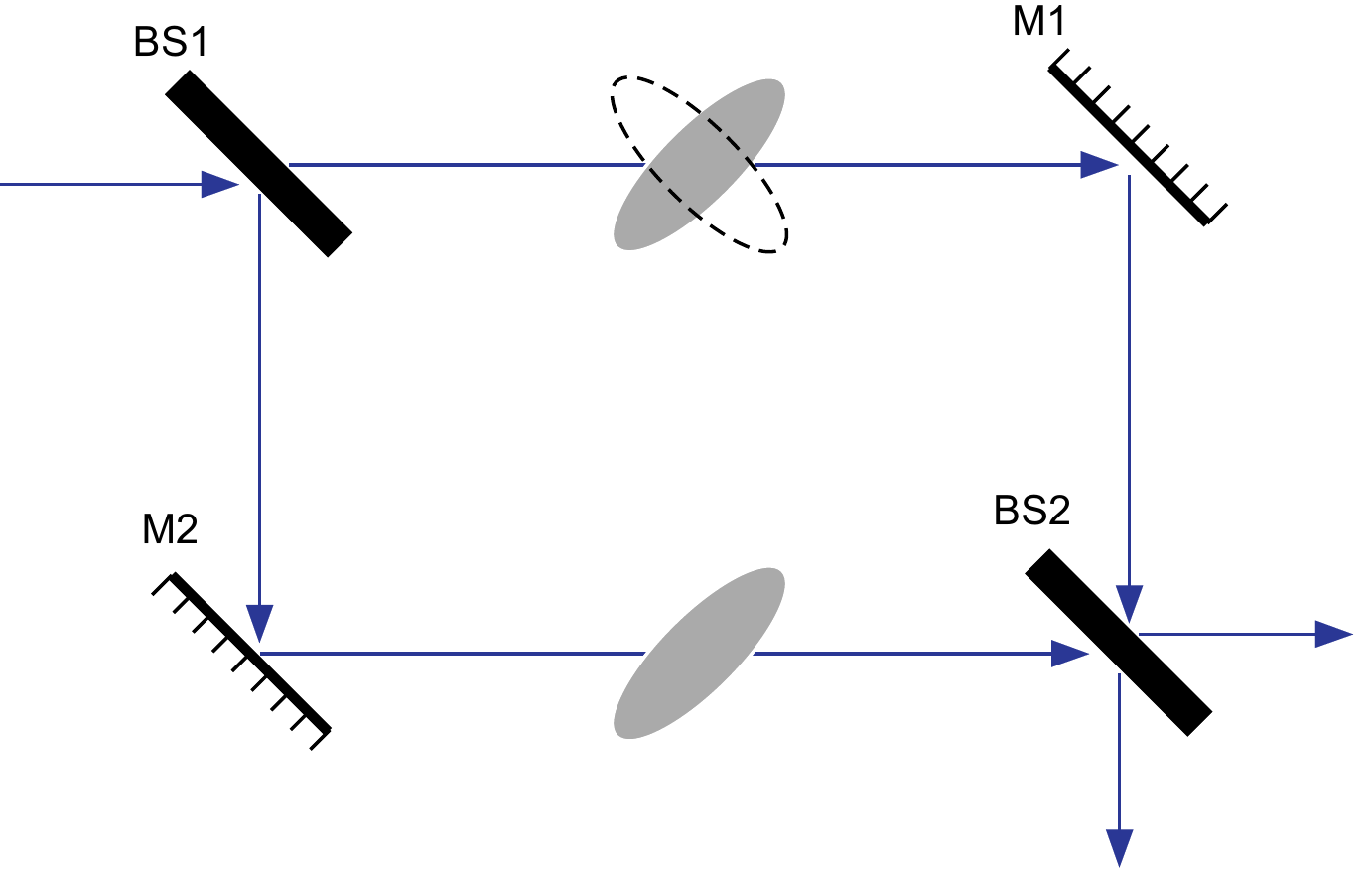}
    \caption{\label{fig:MachZehnder} The evolution due to hamiltonian \eqref{hamiltonian} is equivalent a Mach-Zehnder setup 
    for the resonant case, such that one arm of the interferometer experiences elliptic rotation and the other arm undergoes 
    elliptic rotation (squeezing) if
    $g<g_c$ ($g>g_c$). BS: Beam splitter, M: Mirror.}
\end{figure}

In the resonant case, i.e. $\omega_1 =\omega_2$ and so $A=B=\frac{\pi}{4}$,
the transformation operator $\hat{T}$ in Eq.(\ref{T-op}) becomes exactly the same as the 50:50 beam splitter operator.
Thus the evolution operator $\mathcal{\Op{U}}(t)= e^{-i\Op{H}t}$ corresponding to the hamiltonian (\ref{hamiltonian}) can
be factorized into $\mathcal{\Op{U}}(t) = \Opd{T}e^{-i\op{H}{-}'t}e^{-i\op{H}{+}'t}\Op{T}$, where
$\hat{H}_\pm' = \frac{1}{2}\left(\omega_{1,2}\op{p}{1,2}^2
+(g_c \pm g)\op{x}{1,2}^2\right)$. It is now obvious that in the
resonant case, the dynamics is equivalently represented by a Mach-Zehnder interferometer, with 50:50 beam-splitters as shown
in \fref{fig:MachZehnder}; but with some active operations in both arms \cite{PatKim05}.

To determine the kind of active operations, consider the individual evolutions generated by $\hat{H}_\pm$ which are of the quadratic form,
$\hat{H}_q=\frac{1}{2}(\alpha_q^2 \hat{x}_q^2 + \beta_q^2\hat{p}_q^2)$ ($q=\pm$). The evolution of the quadrature operators induced
by such a hamiltonian is
\begin{equation*}
    \begin{pmatrix}
      \op{x}{q}(t) \\
      \op{p}{q}(t) \\
    \end{pmatrix} = \begin{pmatrix}
                      \cos \alpha_q \beta_q t & -{\beta_q \over \alpha_q}\sin \alpha_q \beta_q t \\
                      {\alpha_q \over \beta_q}\sin \alpha_q \beta_q t & \cos \alpha_q \beta_q t \\
                    \end{pmatrix}
                    \begin{pmatrix}
                        \op{x}{q}(0) \\
                        \op{p}{q}(0) \\
                    \end{pmatrix}.
\end{equation*}
The transformation shows the behavior of an elliptical rotator where the product of the quadrature scaling factors is
unity. It preserves the quadrature uncertainties, leading to at most, squeezing of the mode. Importantly,
if either $\alpha_q$ or $\beta_q$ is negative, the elliptical rotator becomes an elliptical squeezer -- this is
precisely the situation for the `$-$' mode when the coupling is ultra strong, i.e. $g>g_c$.
It is also worth noting that in the general non-resonant case, both the
beam-splitting and two-mode squeezing terms appear in $\Op{T}$, meaning that instead of the
passive beam-splitters at the entry and exit ports, one will have another active device in the equivalent picture.

From the equivalent Mach-Zehnder picture, the source of entanglement is obvious -- squeezing in the input fields is a necessary 
condition for Gaussian entangled output from a beam splitter \cite{KimSon02}. The interaction leads to significantly more entanglement 
when $g=g_c$, than the case where $g<g_c$. To explicitly study dynamic entanglement, we consider each harmonic oscillator initially in 
its thermal equilibrium so that the total density operator
$\Op{\rho}(0)=\op{\rho}{1}^\text{Th}\otimes \op{\rho}{2}^\text{Th}$, where
$ \op{\rho}{j}^\text{Th} = \frac{1}{\mathcal{Z}_j}\exp(-{\delta_j\over 2} (\hat{x}_j^2+\hat{p}_j^2))$ with the canonical partition function
$\mathcal{Z}_j$ and mean excitation number $\eta_j=(\mathrm{e}^{\delta_j}-1)^{-1}$. $\delta_j=\omega_j/k_BT_j$ at temperature $T_j$ with
the Boltzmann constant $k_B$.

\begin{figure}[h!]
    \centering
    \includegraphics[scale=0.4]{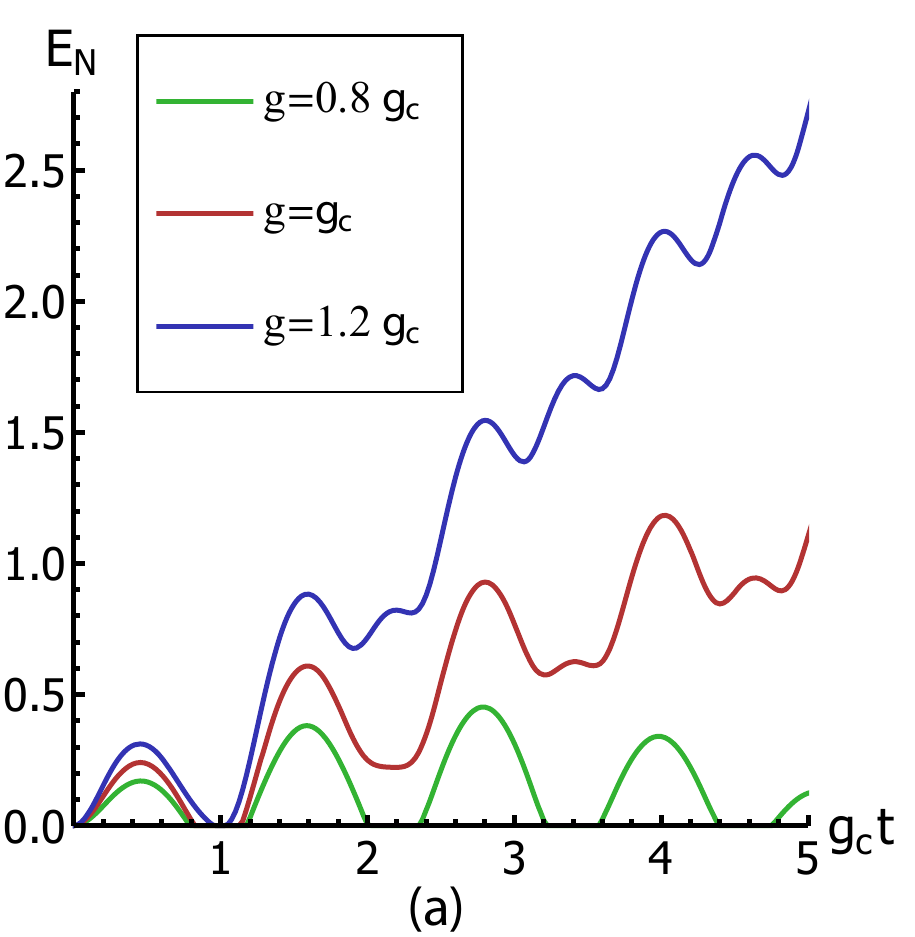}
    \includegraphics[scale=0.4]{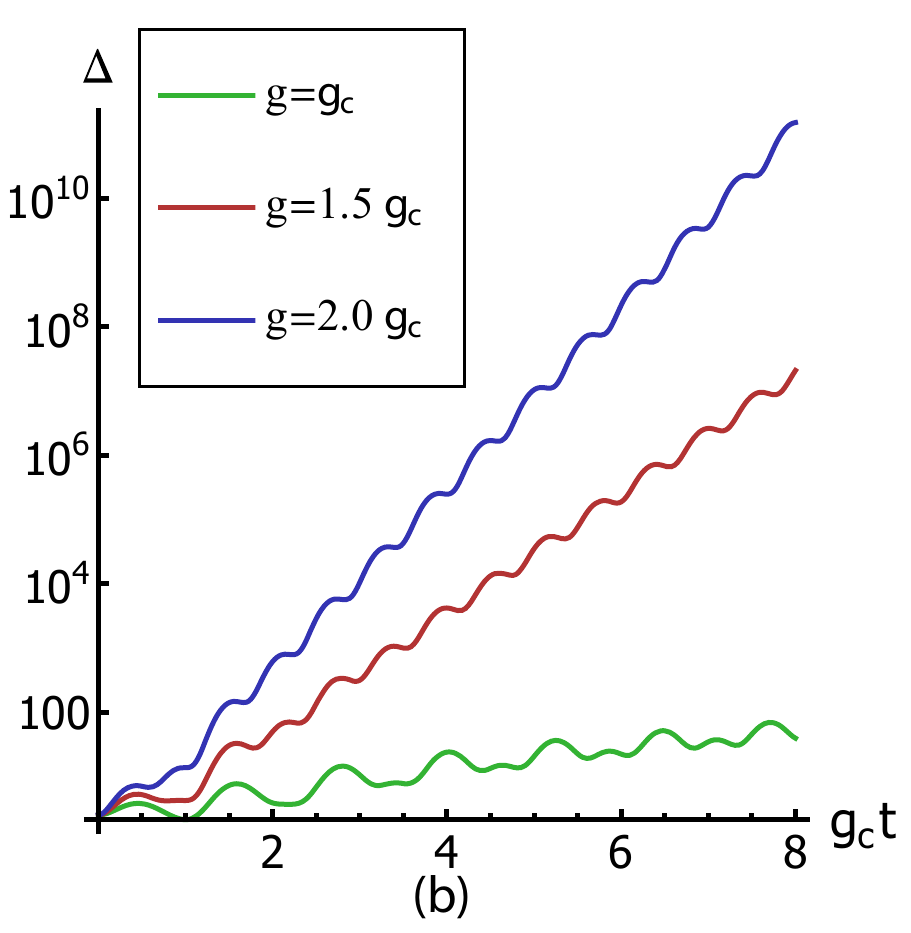}
    \caption{\label{fig:logNeg1}(Color online) (a) Log negativity, and (b) the seralian, for the thermal state with
    $\eta_1=0, \eta_2=1$ for various coupling strengths, at $\omega_1=5\omega_2$.}
\end{figure}

\begin{figure}[h!]
    \centering
    \includegraphics[scale=0.4]{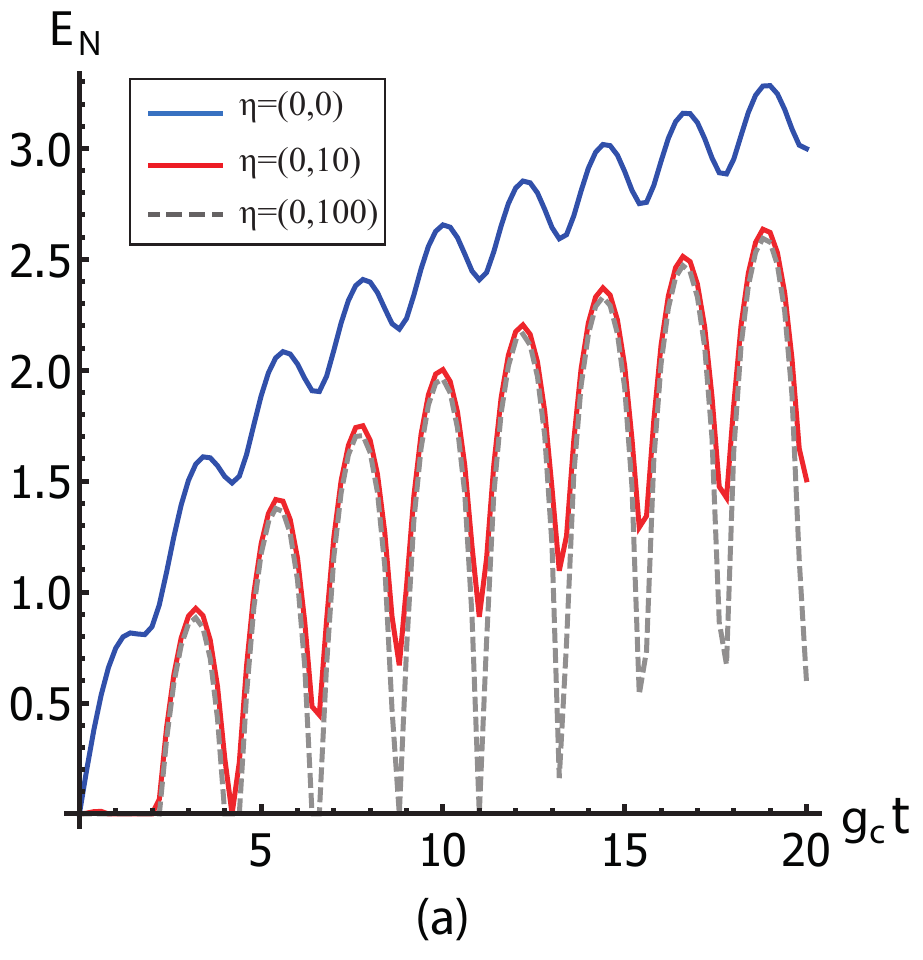}
    \includegraphics[scale=0.4]{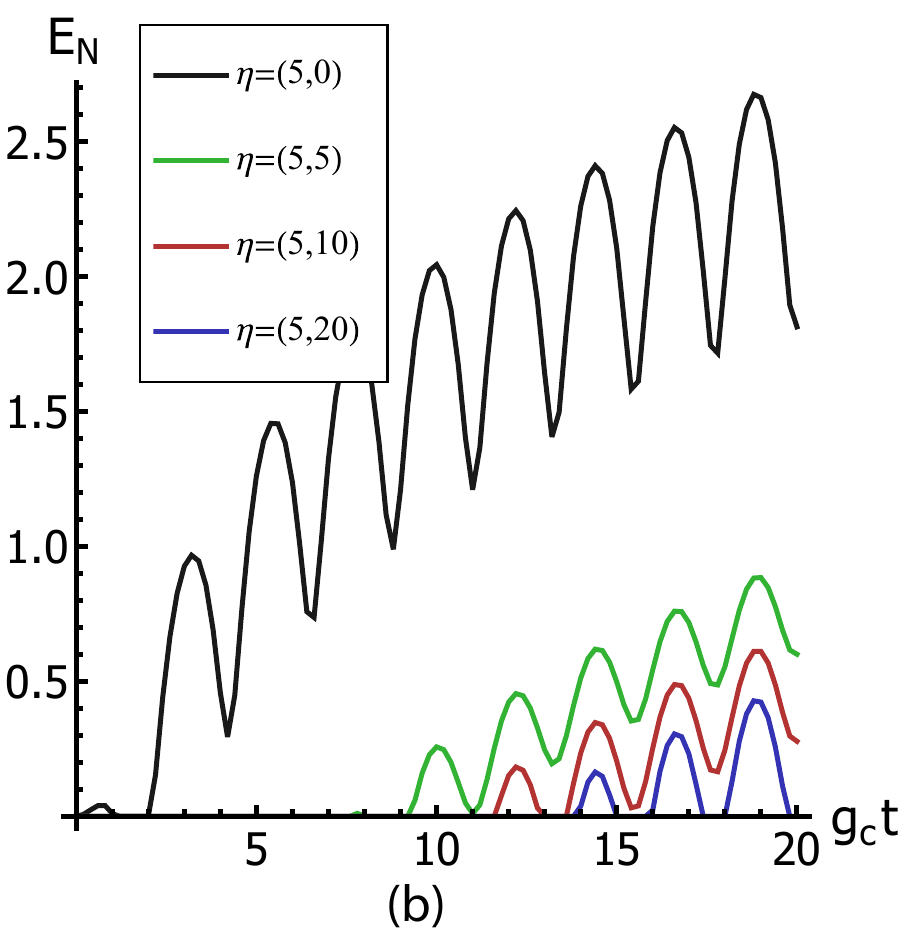}
    \caption{\label{fig:logNeg2}(Color online) Dynamic entanglement at $g=g_c$ for increasing degree of thermality when one oscillator
    is initially (a) in a pure state ($\eta_1=0$) and (b) in a thermally excited state ($\eta_1=5$). The two oscillators are resonant.}
\end{figure}

\fref{fig:logNeg1}(a) shows the dynamic entanglement, measured by the logarithmic negativity \cite{VidWer02,*LeeKim00,Plen05},
for a thermal state with a
single average excitation (see the appendix \ref{app:logneg} for explicit formulas
of the log-negativity). We observe that for $g<g_c$, the entanglement has an oscillatory
behaviour, with typical periods related to the frequencies of the normal modes.
When $g=g_c$, the entanglement monotonically increases, disregarding the small oscillations
due to the `$+$' normal mode. The same behaviour is observed for larger values of the
coupling constant ($g>g_c$) in \fref{fig:logNeg1}(b), where we plot the ``seralian'' $\Delta$ \cite{seralian} 
which, in the unitary case, is a sufficient
entanglement monotone (see Appendix \ref{app:logneg} for details). The entanglement keeps
increasing with time (neglecting the small
oscillations), and in particular at a fixed time, the larger is the value of the coupling
constant $g$, the larger is the entanglement achieved. This particular behaviour
can be explained also by looking at the eigenvalues of the Hamiltonian matrix $H$.
For $g<g_c$, the matrix is positive definite and, as proved in \cite{Gen11}, the corresponding
unitary operator will recur to the identity for a given time; for larger values of $g$, the
matrix $H$ is not positive definite anymore, the recurrence property of the
evolution is lost and the entanglement increases with time.

Even when the initial state is significantly thermal, if the system is critically coupled, entanglement can be generated, 
as depicted in \fref{fig:logNeg2}. In fact, we also observe that the purity of one oscillator is an important factor in the 
generation of entanglement. When one oscillator is pure, the maximum degree
of entanglement achieved remains about the same regardless of the temperature of the other
oscillator. The importance of single-system purity was also observed in some other interaction models \cite{BoseKnight01,KimKnight02}

\subsection{Dissipative dynamics}
Let us consider now the case where the system interacts with a noisy environment.
Because of the environment interaction we observe that also super-critical regime $g>g_c$ can stop to be dynamically unstable.
Assuming that each bare oscillator interacts with its respective thermal environment with mean excitation $\bar{n}_j$, under 
the Born-Markov approximation, the evolution of the system density matrix is governed by the Kossakowski-Lindblad master equation,
\begin{equation}
\label{master}
    \frac{d\Op{\rho}}{dt} = -i[\Op{H},\Op{\rho}] + \sum_{j=1}^2 \frac{\gamma_j}{2} \left\{
        (\bar{n}_j +1)\mathcal{L}[\op{a}{j}] +\bar{n}_j \mathcal{L}[\opd{a}{j}] \right\} \Op{\rho},
\end{equation}
where $\mathcal{L}[\Op{O}]\Op{\rho} = 2\Op{O}\Op{\rho}\Opd{O}-\Opd{O}\Op{O}\Op{\rho}
-\Op{\rho}\Opd{O}\Op{O}$ is the Liouvillian, while $\gamma_j$ is the rate at which the $j^\text{th}$ mode is coupled
to the environment (see Appendix \ref{app:ME} for the analytical solution of \eqref{master})

\begin{figure}[h]
    \centering
    \includegraphics[scale=0.75]{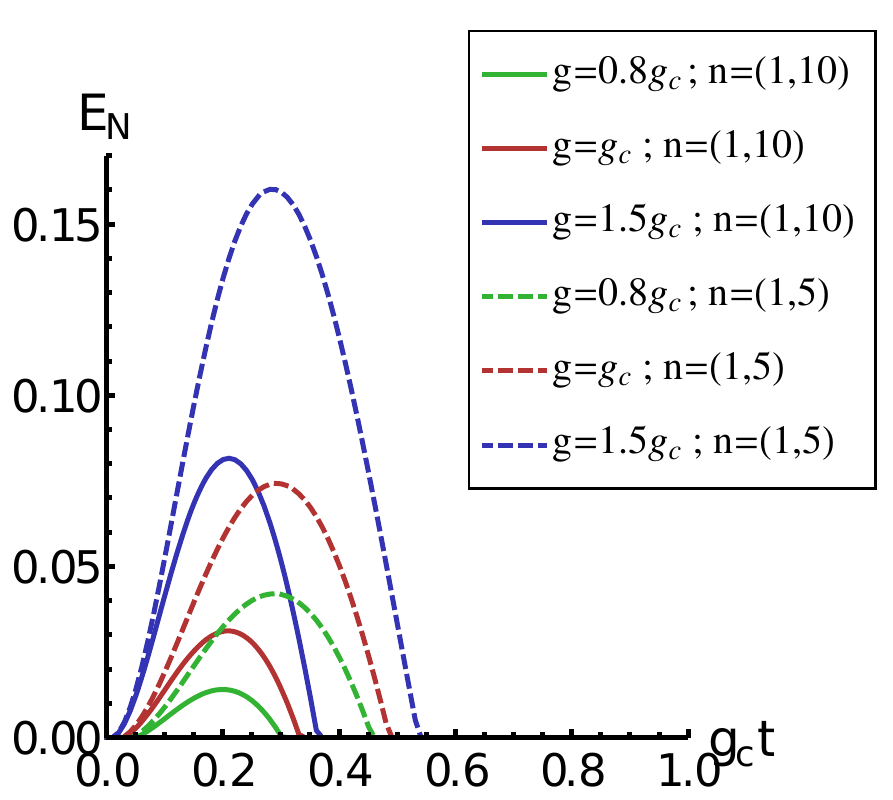}
    \caption{\label{fig:logNegDiss} (Color online) Dynamic entanglement for the dissipative case, for each of the three regimes.
    Here the oscillator $1$ ($2$) is coupled to an environment of $\bar{n}_1 =1$ at a rate $\gamma_1 = 0.01\omega_1$ 
    ($\gamma_2 =0.25 \omega_2$).}
\end{figure}
The logarithmic negativity $E_N$ can be evaluated, and the results are plotted in
\fref{fig:logNegDiss} for various values of the coupling strenght.
We observe that, as expected, also in the super-critical regime
$g>g_c$, the entanglement reaches a maximum value and then starts to
decrease and eventually reaches zero for a given time $t_0$. Not surprisingly, larger
values of the coupling constant and lower values of the noisy parameters
$\eta_j$, $\gamma_j$ and $\bar{n}_j$, correspond to larger maximum values of
entanglement and to larger values of the zero-entanglement time $t_0$.
\section{Conclusion}
We have investigated the ultra-strong-coupled regime of a linear $x-x$ coupled quantum harmonic oscillator
system. Three different regimes of qualitatively different dynamical behaviour can be identified from
the normal modes of the system.
In particular, in the super-critical regime $g>g_c$, the `$-$' mode has an anomalous accelerating motion, which can be
understood as a consequence of a lack of translation symmetry in
the hamiltonian \eqref{hamiltonian}, so that
momentum conservation is not explicitly guaranteed. In
the classical Hookian hamiltonian \eqref{hamiltonian_classical} such a symmetry
is manifest, so that no dynamical anomaly is allowed in the strong coupling regime.

We further find that these three regimes characterized by their dynamical behaviour, is also uniquely characterized
by the dynamics of entanglement; in particular, that the super-critical regime is one where
entanglement can be unboundedly generated from even a highly thermal state.
A similar behavior was studied in the context of the superradiant phase transition of the Dicke model, whose
effective hamiltonian is of the form \eqref{hamiltonian} after a Holstein-Primakoff transformation of a set
of SU(2) operators into a bosonic operator.

Finally, we also note that the super-critical regime exhibits a finite entanglement even though
the system maybe coupled to a decohering Markovian environment at a non-zero temperature.

\begin{acknowledgements}
MSK thanks Wonmin Son for discussions. This work was supported by the Qatar National Research Fund (NPRP 4-554-1-084). 
JL acknowledges support from a National Research Foundation of Korea (NRF) grant (No. 2010-0015059). MGG acknolwedges a 
fellowship support from UK EPSRC (EP/I026436/1) and thanks Alessio Serafini and Matteo Binazzi for interesting and useful 
discussions.
\end{acknowledgements}

\appendix
\section{Log-negativity for Gaussian states} \label{app:logneg}
Given a two-mode Gaussian state, its entanglement properties are
fully characterized by its covariance matrix $\sig$, which can be written
in terms of $2\times 2$ matrices as
\begin{align}
\sig =
\left(
\begin{array}{c c}
\sig_1 &  \boldsymbol{\gamma} \\
\boldsymbol{\gamma}^\trans & \sig_2 \\
\end{array}
\right)\, .
\end{align}
The log-negativity \cite{VidWer02} of the two-mode state can be evaluated
as $E_N(\varrho) = \max[0,-\log(2\tilde{ \nu}_{-})]$, where $\tilde{\nu}_{-}$
is the lowest symplectic eigenvalue of the corresponding partial transposed state.
In formula we have
\begin{align}
\tilde{\nu}_-^2 = \frac{\Delta - \sqrt{\Delta^2 - 4 \,{\rm Det}[\sig]}}{2}
\end{align}
where the ``seralian'' $\Delta$ is defined as
\begin{align}
\Delta = {\rm Det}[\sig_1] + {\rm Det}[\sig_2] - 2 {\rm Det}[\boldsymbol{\gamma}]
\end{align}
and where ${\rm Det}[A]$ denotes the determinant of the matrix $A$. It is known that the
logarithmic negativity $E_N$ is an entanglement monotone \cite{Plen05}. One should also note
that ${\rm Det}[\sig]$ is directly related to the purity of the state, since $\Tr[\varrho^2]=1/(2 \sqrt{{\rm Det}[\sig]})$.
As a consequence, unitary evolution leaves ${\rm Det}[\sig]$ invariant, so that the log-negativity is monotonous
with $\Delta$. For this reason, when we consider the entanglement dynamics due to non-dissipative evolution,
we may take the seralian $\Delta$ to be an entanglement monotone.
\section{Solving the master equation}\label{app:ME}
The master equation \eqref{master} can be cast into a c-number partial differential equation (PDE) by introducing the symmetric ordered
characteristic function $$\chi (\alpha_1,\alpha_2;t)=\Tr[\op{D}{1}(\alpha_1)\op{D}{2}(\alpha_2)\Op{\rho}(t)]\, ,$$ where
$\op{D}{j}(\alpha_j)=\exp(\alpha_j \opd{a}{j}-\alpha_j^* \op{a}{j})$. We choose
to express the PDE in terms of the quadratures parameters $x_j,p_j$ defined via $\alpha_{j}=\frac{1}{\sqrt{2}}(x_{j}+i{p_j})$, so that
we get the Fokker-Planck equation,
\begin{equation}\label{FPE}
\begin{split}
    \frac{\partial \chi}{\partial t} = -\frac{1}{2}({\sf r}^\trans \bar{\Gamma} {\sf r})\chi
        & +\frac{1}{2} \left[{\sf r}^\trans \left(\Upsilon \widetilde{H}-\frac{1}{2}\Gamma\right)
        \overrightarrow{\partial_{\sf r}}\right]\chi \\
        & +\frac{1}{2} \chi \left[\overleftarrow{\partial_{\sf r}^\trans} \left((\Upsilon \widetilde{H})^\trans-\frac{1}{2}\Gamma\right)
        {\sf r}\right],
\end{split}
\end{equation}
where we define the vector ${\sf r}=(x_1,p_1,x_2,p_2)^\trans$, the associated gradient operator
$\partial_{\sf r} =(\partial_{x_1},\partial_{p_1},\partial_{x_2},\partial_{p_2})^\trans $, the following matrices:
\begin{align*}
    \Upsilon &= \bigoplus_{j=1}^2 \begin{bmatrix} 0 & 1 \\ -1 & 0 \end{bmatrix}, \,\,\,\,
    \Gamma = \bigoplus_{j=1}^2 \begin{bmatrix} \gamma_j & 0 \\ 0 & \gamma_j \end{bmatrix}, \\
    \bar{\Gamma} &= \bigoplus_{j=1}^2 \begin{bmatrix} \gamma_j \left(\bar{n}_j +\frac{1}{2} \right)
    & 0 \\     0 & \gamma_j \left(\bar{n}_j +\frac{1}{2} \right) \end{bmatrix},
\end{align*}
and where $\widetilde{H}$ here is a permutation on the hamiltonian matrix \eqref{hamiltonian_matrix} corresponding to the 
difference in the order
chosen for the quadrature variables. The arrows on the gradient operators denote the direction in which
the differential operators act.

As shown above, the system Hamiltonian only affects elliptic rotations or elliptic squeezing in the course of
the dynamics, so that an initial Gaussian state localized at the origin in phase space remains so always; and since the bath is in thermal
equilibrium, it too does not affect any finite displacements in phase space, but only scale and rotation changes. Thus, for our
relevant case, without loss of generality, we choose the ansatz,
\begin{equation*}
    \chi ({\sf r};t) = \exp\left(-\frac{1}{2}{\sf r}^\trans \sigma(t) {\sf r} \right),
\end{equation*}
where now the time dependence is carried in the covariance matrix. Substituting this into \eqref{FPE}, and then identifying 
the coefficients of the various bilinear products, we get the equation of motion for the covariance matrix viz.,
\begin{equation*}
    \frac{d\sigma}{dt} + \left(\frac{1}{2}\Gamma - \Upsilon\widetilde{H} \right)\sigma
                +\sigma \left(\frac{1}{2}\Gamma - \Upsilon\widetilde{H}\right)^\trans = \bar{\Gamma}.
\end{equation*}
This has the solution,
\begin{equation*}
    \sigma(t) = K(t) \left\{\sigma(0) +\int_0^t K(-\tau) \bar{\Gamma} K^\trans (-\tau)\, d\tau  \right\} K^\trans(t),
\end{equation*}
where $K(t) = \exp\left[ \left( \Upsilon\widetilde{H}-\frac{1}{2}\Gamma \right)t\right]$.

\bibliography{sc_paper_refs}

\begin{thebibliography}{30}
\expandafter\ifx\csname natexlab\endcsname\relax\def\natexlab#1{#1}\fi
\expandafter\ifx\csname bibnamefont\endcsname\relax
  \def\bibnamefont#1{#1}\fi
\expandafter\ifx\csname bibfnamefont\endcsname\relax
  \def\bibfnamefont#1{#1}\fi
\expandafter\ifx\csname citenamefont\endcsname\relax
  \def\citenamefont#1{#1}\fi
\expandafter\ifx\csname url\endcsname\relax
  \def\url#1{\texttt{#1}}\fi
\expandafter\ifx\csname urlprefix\endcsname\relax\def\urlprefix{URL }\fi
\providecommand{\bibinfo}[2]{#2}
\providecommand{\eprint}[2][]{\url{#2}}

\bibitem[{\citenamefont{Autler and Townes}(1955)}]{AutTow55}
\bibinfo{author}{\bibfnamefont{S.}~\bibnamefont{Autler}} \bibnamefont{and}
  \bibinfo{author}{\bibfnamefont{C.}~\bibnamefont{Townes}},
  \bibinfo{journal}{Phys. Rev.} \textbf{\bibinfo{volume}{100}},
  \bibinfo{pages}{703} (\bibinfo{year}{1955}).

\bibitem[{\citenamefont{Raimond et~al.}(2001)\citenamefont{Raimond, Brune, and
  Haroche}}]{RaiHar01}
\bibinfo{author}{\bibfnamefont{J.~M.} \bibnamefont{Raimond}},
  \bibinfo{author}{\bibfnamefont{M.}~\bibnamefont{Brune}}, \bibnamefont{and}
  \bibinfo{author}{\bibfnamefont{S.}~\bibnamefont{Haroche}},
  \bibinfo{journal}{Rev. Mod. Phys.} \textbf{\bibinfo{volume}{73}},
  \bibinfo{pages}{565} (\bibinfo{year}{2001}).

\bibitem[{\citenamefont{Haroche and Raimond}(2006)}]{HarBook}
\bibinfo{author}{\bibfnamefont{S.}~\bibnamefont{Haroche}} \bibnamefont{and}
  \bibinfo{author}{\bibfnamefont{J.~M.} \bibnamefont{Raimond}},
  \emph{\bibinfo{title}{Exploring the quantum}} (\bibinfo{publisher}{Oxford
  University Press}, \bibinfo{year}{2006}).

\bibitem[{\citenamefont{Verhagen et~al.}(2012)\citenamefont{Verhagen,
  Deleglise, Weis, Schliesser, and Kippenberg}}]{Verh12}
\bibinfo{author}{\bibfnamefont{E.}~\bibnamefont{Verhagen}},
  \bibinfo{author}{\bibfnamefont{S.}~\bibnamefont{Deleglise}},
  \bibinfo{author}{\bibfnamefont{S.}~\bibnamefont{Weis}},
  \bibinfo{author}{\bibfnamefont{A.}~\bibnamefont{Schliesser}},
  \bibnamefont{and} \bibinfo{author}{\bibfnamefont{T.~J.}
  \bibnamefont{Kippenberg}}, \bibinfo{journal}{Nature}
  \textbf{\bibinfo{volume}{482}}, \bibinfo{pages}{63} (\bibinfo{year}{2012}).

\bibitem[{\citenamefont{Wallraff et~al.}(2004)\citenamefont{Wallraff, Schuster,
  Blais, Frunzio, Huang, Majer, Kumar, Girvin, and Schoelkopf}}]{WalScho04}
\bibinfo{author}{\bibfnamefont{A.}~\bibnamefont{Wallraff}},
  \bibinfo{author}{\bibfnamefont{D.~I.} \bibnamefont{Schuster}},
  \bibinfo{author}{\bibfnamefont{A.}~\bibnamefont{Blais}},
  \bibinfo{author}{\bibfnamefont{L.}~\bibnamefont{Frunzio}},
  \bibinfo{author}{\bibfnamefont{R.~S.} \bibnamefont{Huang}},
  \bibinfo{author}{\bibfnamefont{J.}~\bibnamefont{Majer}},
  \bibinfo{author}{\bibfnamefont{S.}~\bibnamefont{Kumar}},
  \bibinfo{author}{\bibfnamefont{S.~M.} \bibnamefont{Girvin}},
  \bibnamefont{and} \bibinfo{author}{\bibfnamefont{R.~J.}
  \bibnamefont{Schoelkopf}}, \bibinfo{journal}{Nature}
  \textbf{\bibinfo{volume}{431}}, \bibinfo{pages}{162} (\bibinfo{year}{2004}).

\bibitem[{\citenamefont{Niemczyk et~al.}(2010)\citenamefont{Niemczyk, Deppe,
  Huebl, Menzel, Hocke, Schwarz, Garcia-Ripoo, Zueco, H\"{u}mmer, Solano
  et~al.}}]{NieSol10}
\bibinfo{author}{\bibfnamefont{T.}~\bibnamefont{Niemczyk}},
  \bibinfo{author}{\bibfnamefont{F.}~\bibnamefont{Deppe}},
  \bibinfo{author}{\bibfnamefont{H.}~\bibnamefont{Huebl}},
  \bibinfo{author}{\bibfnamefont{E.}~\bibnamefont{Menzel}},
  \bibinfo{author}{\bibfnamefont{F.}~\bibnamefont{Hocke}},
  \bibinfo{author}{\bibfnamefont{M.}~\bibnamefont{Schwarz}},
  \bibinfo{author}{\bibfnamefont{J.}~\bibnamefont{Garcia-Ripoo}},
  \bibinfo{author}{\bibfnamefont{D.}~\bibnamefont{Zueco}},
  \bibinfo{author}{\bibfnamefont{T.}~\bibnamefont{H\"{u}mmer}},
  \bibinfo{author}{\bibfnamefont{E.}~\bibnamefont{Solano}},
  \bibnamefont{et~al.}, \bibinfo{journal}{Nature Phys.}
  \textbf{\bibinfo{volume}{6}}, \bibinfo{pages}{772} (\bibinfo{year}{2010}).

\bibitem[{\citenamefont{Casanova et~al.}(2010)\citenamefont{Casanova, Romero,
  Lizuain, Garcia-Ripoll, and Solano}}]{CasSol10}
\bibinfo{author}{\bibfnamefont{J.}~\bibnamefont{Casanova}},
  \bibinfo{author}{\bibfnamefont{G.}~\bibnamefont{Romero}},
  \bibinfo{author}{\bibfnamefont{I.}~\bibnamefont{Lizuain}},
  \bibinfo{author}{\bibfnamefont{J.~J.} \bibnamefont{Garcia-Ripoll}},
  \bibnamefont{and} \bibinfo{author}{\bibfnamefont{E.}~\bibnamefont{Solano}},
  \bibinfo{journal}{Phys. Rev. Lett.} \textbf{\bibinfo{volume}{105}},
  \bibinfo{pages}{263603} (\bibinfo{year}{2010}).

\bibitem[{\citenamefont{Hopfield}(1958)}]{Hop58}
\bibinfo{author}{\bibfnamefont{J.~J.} \bibnamefont{Hopfield}},
  \bibinfo{journal}{Phys. Rev.} \textbf{\bibinfo{volume}{112}},
  \bibinfo{pages}{1555} (\bibinfo{year}{1958}).

\bibitem[{\citenamefont{Eisert et~al.}(2004)\citenamefont{Eisert, Plenio, Bose,
  and Hartley}}]{EisPlenBos04}
\bibinfo{author}{\bibfnamefont{J.}~\bibnamefont{Eisert}},
  \bibinfo{author}{\bibfnamefont{M.~B.} \bibnamefont{Plenio}},
  \bibinfo{author}{\bibfnamefont{S.}~\bibnamefont{Bose}}, \bibnamefont{and}
  \bibinfo{author}{\bibfnamefont{J.}~\bibnamefont{Hartley}},
  \bibinfo{journal}{Phys. Rev. Lett.} \textbf{\bibinfo{volume}{93}},
  \bibinfo{pages}{190402} (\bibinfo{year}{2004}).

\bibitem[{\citenamefont{Audenaert et~al.}(2002)\citenamefont{Audenaert, Eisert,
  Plenio, and Werner}}]{AudPlenWer02}
\bibinfo{author}{\bibfnamefont{J.~K.} \bibnamefont{Audenaert}},
  \bibinfo{author}{\bibfnamefont{J.}~\bibnamefont{Eisert}},
  \bibinfo{author}{\bibfnamefont{M.~B.} \bibnamefont{Plenio}},
  \bibnamefont{and} \bibinfo{author}{\bibfnamefont{R.~F.}
  \bibnamefont{Werner}}, \bibinfo{journal}{Phys. Rev. A}
  \textbf{\bibinfo{volume}{66}}, \bibinfo{pages}{042327}
  (\bibinfo{year}{2002}).

\bibitem[{\citenamefont{Galve et~al.}(2010)\citenamefont{Galve, Pachon, and
  Zueco}}]{Galv10}
\bibinfo{author}{\bibfnamefont{F.}~\bibnamefont{Galve}},
  \bibinfo{author}{\bibfnamefont{L.~A.} \bibnamefont{Pachon}},
  \bibnamefont{and} \bibinfo{author}{\bibfnamefont{D.}~\bibnamefont{Zueco}},
  \bibinfo{journal}{Phys. Rev. Lett.} \textbf{\bibinfo{volume}{105}},
  \bibinfo{pages}{180501} (\bibinfo{year}{2010}).

\bibitem[{\citenamefont{Ciuti et~al.}(2005)\citenamefont{Ciuti, Bastard, and
  Carusotto}}]{CiuCar05}
\bibinfo{author}{\bibfnamefont{C.}~\bibnamefont{Ciuti}},
  \bibinfo{author}{\bibfnamefont{G.}~\bibnamefont{Bastard}}, \bibnamefont{and}
  \bibinfo{author}{\bibfnamefont{I.}~\bibnamefont{Carusotto}},
  \bibinfo{journal}{Phys. Rev. B} \textbf{\bibinfo{volume}{72}},
  \bibinfo{pages}{115303} (\bibinfo{year}{2005}).

\bibitem[{\citenamefont{G\"{u}nter et~al.}(2009)\citenamefont{G\"{u}nter,
  Anappara, Hees, Sell, Biasiol, Sorba, Liberato, Ciuti, Tredicucci,
  Leitenstorfer et~al.}}]{GunCiuHub05}
\bibinfo{author}{\bibfnamefont{G.}~\bibnamefont{G\"{u}nter}},
  \bibinfo{author}{\bibfnamefont{A.~A.} \bibnamefont{Anappara}},
  \bibinfo{author}{\bibfnamefont{J.}~\bibnamefont{Hees}},
  \bibinfo{author}{\bibfnamefont{A.}~\bibnamefont{Sell}},
  \bibinfo{author}{\bibfnamefont{G.}~\bibnamefont{Biasiol}},
  \bibinfo{author}{\bibfnamefont{L.}~\bibnamefont{Sorba}},
  \bibinfo{author}{\bibfnamefont{S.~D.} \bibnamefont{Liberato}},
  \bibinfo{author}{\bibfnamefont{C.}~\bibnamefont{Ciuti}},
  \bibinfo{author}{\bibfnamefont{A.}~\bibnamefont{Tredicucci}},
  \bibinfo{author}{\bibfnamefont{A.}~\bibnamefont{Leitenstorfer}},
  \bibnamefont{et~al.}, \bibinfo{journal}{Nature}
  \textbf{\bibinfo{volume}{458}}, \bibinfo{pages}{178} (\bibinfo{year}{2009}).

\bibitem[{\citenamefont{Bina et~al.}(2012)\citenamefont{Bina, Romero, Casanova,
  Garcia-Ripoll, Lulli, Casagrande, and Solano}}]{bina11}
\bibinfo{author}{\bibfnamefont{M.}~\bibnamefont{Bina}},
  \bibinfo{author}{\bibfnamefont{G.}~\bibnamefont{Romero}},
  \bibinfo{author}{\bibfnamefont{J.}~\bibnamefont{Casanova}},
  \bibinfo{author}{\bibfnamefont{J.~J.} \bibnamefont{Garcia-Ripoll}},
  \bibinfo{author}{\bibfnamefont{A.}~\bibnamefont{Lulli}},
  \bibinfo{author}{\bibfnamefont{F.}~\bibnamefont{Casagrande}},
  \bibnamefont{and} \bibinfo{author}{\bibfnamefont{E.}~\bibnamefont{Solano}},
  \bibinfo{journal}{Eur. Phys. J.} \textbf{\bibinfo{volume}{203}},
  \bibinfo{pages}{207} (\bibinfo{year}{2012}).

\bibitem[{\citenamefont{Dicke}(1968)}]{Dicke68}
\bibinfo{author}{\bibfnamefont{R.~H.} \bibnamefont{Dicke}},
  \bibinfo{journal}{Phys. Rev.} \textbf{\bibinfo{volume}{93}},
  \bibinfo{pages}{99} (\bibinfo{year}{1968}).

\bibitem[{\citenamefont{Mancini and Tombesi}(1994)}]{MancTomb94}
\bibinfo{author}{\bibfnamefont{S.}~\bibnamefont{Mancini}} \bibnamefont{and}
  \bibinfo{author}{\bibfnamefont{P.}~\bibnamefont{Tombesi}},
  \bibinfo{journal}{Phys. Rev. A} \textbf{\bibinfo{volume}{49}},
  \bibinfo{pages}{4055} (\bibinfo{year}{1994}).

\bibitem[{\citenamefont{Akram et~al.}(2010)\citenamefont{Akram, Kiesel,
  Aspelmeyer, and Milburn}}]{AkrAspMil10}
\bibinfo{author}{\bibfnamefont{U.}~\bibnamefont{Akram}},
  \bibinfo{author}{\bibfnamefont{N.}~\bibnamefont{Kiesel}},
  \bibinfo{author}{\bibfnamefont{M.}~\bibnamefont{Aspelmeyer}},
  \bibnamefont{and} \bibinfo{author}{\bibfnamefont{G.~J.}
  \bibnamefont{Milburn}}, \bibinfo{journal}{New J. Phys.}
  \textbf{\bibinfo{volume}{12}}, \bibinfo{pages}{083030}
  (\bibinfo{year}{2010}).

\bibitem[{\citenamefont{McDuff and Salamon}(1998)}]{DufSal98}
\bibinfo{author}{\bibfnamefont{D.}~\bibnamefont{McDuff}} \bibnamefont{and}
  \bibinfo{author}{\bibfnamefont{D.}~\bibnamefont{Salamon}},
  \emph{\bibinfo{title}{Introduction to Symplectic Topology}}
  (\bibinfo{publisher}{Oxford University Press}, \bibinfo{year}{1998}).

\bibitem[{\citenamefont{Law}(1995)}]{Law94}
\bibinfo{author}{\bibfnamefont{C.~K.} \bibnamefont{Law}},
  \bibinfo{journal}{Phys. Rev. A} \textbf{\bibinfo{volume}{51}},
  \bibinfo{pages}{2537} (\bibinfo{year}{1995}).

\bibitem[{\citenamefont{Pace et~al.}(1993)\citenamefont{Pace, Collett, and
  Walls}}]{CollWalls93}
\bibinfo{author}{\bibfnamefont{A.~F.} \bibnamefont{Pace}},
  \bibinfo{author}{\bibfnamefont{M.~J.} \bibnamefont{Collett}},
  \bibnamefont{and} \bibinfo{author}{\bibfnamefont{D.~F.} \bibnamefont{Walls}},
  \bibinfo{journal}{Phys. Rev. A} \textbf{\bibinfo{volume}{47}},
  \bibinfo{pages}{3173} (\bibinfo{year}{1993}).

\bibitem[{\citenamefont{Vanner et~al.}(2011)\citenamefont{Vanner, Pikovski,
  Cole, Kim, Brukner, Hammerer, Milburn, and Aspelmeyer}}]{VanPNAS}
\bibinfo{author}{\bibfnamefont{M.~R.} \bibnamefont{Vanner}},
  \bibinfo{author}{\bibfnamefont{I.}~\bibnamefont{Pikovski}},
  \bibinfo{author}{\bibfnamefont{G.~D.} \bibnamefont{Cole}},
  \bibinfo{author}{\bibfnamefont{M.~S.} \bibnamefont{Kim}},
  \bibinfo{author}{\bibfnamefont{C.}~\bibnamefont{Brukner}},
  \bibinfo{author}{\bibfnamefont{K.}~\bibnamefont{Hammerer}},
  \bibinfo{author}{\bibfnamefont{G.~J.} \bibnamefont{Milburn}},
  \bibnamefont{and}
  \bibinfo{author}{\bibfnamefont{M.}~\bibnamefont{Aspelmeyer}},
  \bibinfo{journal}{Proc. Natl. Acad. Sci. USA} \textbf{\bibinfo{volume}{108}},
  \bibinfo{pages}{16182} (\bibinfo{year}{2011}).

\bibitem[{\citenamefont{Paternostro et~al.}(2005)\citenamefont{Paternostro,
  Kim, Park, and Lee}}]{PatKim05}
\bibinfo{author}{\bibfnamefont{M.}~\bibnamefont{Paternostro}},
  \bibinfo{author}{\bibfnamefont{M.~S.} \bibnamefont{Kim}},
  \bibinfo{author}{\bibfnamefont{E.}~\bibnamefont{Park}}, \bibnamefont{and}
  \bibinfo{author}{\bibfnamefont{J.}~\bibnamefont{Lee}},
  \bibinfo{journal}{Phys. Rev. A} \textbf{\bibinfo{volume}{72}},
  \bibinfo{pages}{052307} (\bibinfo{year}{2005}).

\bibitem[{\citenamefont{Kim et~al.}(2002{\natexlab{a}})\citenamefont{Kim, Son,
  Bu\v{z}ek, and Knight}}]{KimSon02}
\bibinfo{author}{\bibfnamefont{M.~S.} \bibnamefont{Kim}},
  \bibinfo{author}{\bibfnamefont{W.}~\bibnamefont{Son}},
  \bibinfo{author}{\bibfnamefont{V.}~\bibnamefont{Bu\v{z}ek}},
  \bibnamefont{and} \bibinfo{author}{\bibfnamefont{P.~L.}
  \bibnamefont{Knight}}, \bibinfo{journal}{Phys. Rev. A}
  \textbf{\bibinfo{volume}{65}}, \bibinfo{pages}{032323}
  (\bibinfo{year}{2002}{\natexlab{a}}).

\bibitem[{\citenamefont{Vidal and Werner}(2002)}]{VidWer02}
\bibinfo{author}{\bibfnamefont{G.}~\bibnamefont{Vidal}} \bibnamefont{and}
  \bibinfo{author}{\bibfnamefont{R.~F.} \bibnamefont{Werner}},
  \bibinfo{journal}{Phys. Rev. A} \textbf{\bibinfo{volume}{65}},
  \bibinfo{pages}{032314} (\bibinfo{year}{2002}).

\bibitem[{\citenamefont{Lee et~al.}(2000)\citenamefont{Lee, Kim, Park, and
  Lee}}]{LeeKim00}
\bibinfo{author}{\bibfnamefont{J.}~\bibnamefont{Lee}},
  \bibinfo{author}{\bibfnamefont{M.~S.} \bibnamefont{Kim}},
  \bibinfo{author}{\bibfnamefont{Y.~J.} \bibnamefont{Park}}, \bibnamefont{and}
  \bibinfo{author}{\bibfnamefont{S.}~\bibnamefont{Lee}}, \bibinfo{journal}{J.
  Mod. Opt.} \textbf{\bibinfo{volume}{47}}, \bibinfo{pages}{2151}
  (\bibinfo{year}{2000}).

\bibitem[{\citenamefont{Plenio}(2005)}]{Plen05}
\bibinfo{author}{\bibfnamefont{M.~B.} \bibnamefont{Plenio}},
  \bibinfo{journal}{Phys. Rev. Lett.} \textbf{\bibinfo{volume}{95}},
  \bibinfo{pages}{090503} (\bibinfo{year}{2005}).

\bibitem[{\citenamefont{Serafini et~al.}(2004)\citenamefont{Serafini,
  Illuminati, and De~Siena}}]{seralian}
\bibinfo{author}{\bibfnamefont{A.}~\bibnamefont{Serafini}},
  \bibinfo{author}{\bibfnamefont{F.}~\bibnamefont{Illuminati}},
  \bibnamefont{and} \bibinfo{author}{\bibfnamefont{S.}~\bibnamefont{De~Siena}},
  \bibinfo{journal}{J. Phys. B} \textbf{\bibinfo{volume}{37}},
  \bibinfo{pages}{L21} (\bibinfo{year}{2004}).

\bibitem[{\citenamefont{Genoni et~al.}(2012)\citenamefont{Genoni, Serafini,
  Kim, and Burgarth}}]{Gen11}
\bibinfo{author}{\bibfnamefont{M.}~\bibnamefont{Genoni}},
  \bibinfo{author}{\bibfnamefont{A.}~\bibnamefont{Serafini}},
  \bibinfo{author}{\bibfnamefont{M.~S.} \bibnamefont{Kim}}, \bibnamefont{and}
  \bibinfo{author}{\bibfnamefont{D.}~\bibnamefont{Burgarth}},
  \bibinfo{journal}{Phys. Rev. Lett.} \textbf{\bibinfo{volume}{108}},
  \bibinfo{pages}{150501} (\bibinfo{year}{2012}).

\bibitem[{\citenamefont{Bose et~al.}(2001)\citenamefont{Bose, Fuentes-Guridi,
  Knight, and Vedral}}]{BoseKnight01}
\bibinfo{author}{\bibfnamefont{S.}~\bibnamefont{Bose}},
  \bibinfo{author}{\bibfnamefont{I.}~\bibnamefont{Fuentes-Guridi}},
  \bibinfo{author}{\bibfnamefont{P.~L.} \bibnamefont{Knight}},
  \bibnamefont{and} \bibinfo{author}{\bibfnamefont{V.}~\bibnamefont{Vedral}},
  \bibinfo{journal}{Phys. Rev. Lett.} \textbf{\bibinfo{volume}{87}},
  \bibinfo{pages}{050401} (\bibinfo{year}{2001}).

\bibitem[{\citenamefont{Kim et~al.}(2002{\natexlab{b}})\citenamefont{Kim, Lee,
  Ahn, and Knight}}]{KimKnight02}
\bibinfo{author}{\bibfnamefont{M.~S.} \bibnamefont{Kim}},
  \bibinfo{author}{\bibfnamefont{J.}~\bibnamefont{Lee}},
  \bibinfo{author}{\bibfnamefont{D.}~\bibnamefont{Ahn}}, \bibnamefont{and}
  \bibinfo{author}{\bibfnamefont{P.~L.} \bibnamefont{Knight}},
  \bibinfo{journal}{Phys. Rev. A} \textbf{\bibinfo{volume}{65}},
  \bibinfo{pages}{040501} (\bibinfo{year}{2002}{\natexlab{b}}).

\end{thebibliography}
\end{document}